# Orientation-dependent surface composition of *in situ* annealed strontium titanate

Luiz F. Zagonel,[a] Nicholas Barrett,[a]* Olivier Renault,[b] Aude Bailly,[b] Michael Bäurer,[c] Michael Hoffmann,[c] Shao-Ju Shih,[d] and David Cockayne,[d]

The surface composition of polycrystalline niobium-doped strontium titanate ($SrTiO_3$ : Nb) is studied using X-ray photoelectron emission microscopy (XPEEM) for many grain orientations in order to characterise the surface chemistry with high spatial resolution. The surface sensitivity is maximised by the use of soft X-ray synchrotron radiation (SR). The grain orientation is determined by electron backscattering diffraction (EBSD). Stereographic plots are used to show the correlation between surface composition and orientation for several grains. Predominant surface terminations are assigned to major orientations.

## Introduction

Strontium titanate (STO) belongs to the family of perovskite oxides of considerable importance for a variety of technological applications. It is widely used as a substrate for thin film deposition[1–3] for ferroelectricity, for epitaxial growth of high $T_C$ superconductors, and as ceramic capacitors for the microelectronics industry. The surface structure can be rather complex and many studies, both theoretical and experimental, have contributed to its understanding. The STO perovskite structure gives alternating $TiO_2$ and SrO planes in the (001) direction. Special attention has been paid to the three main surface planes, (100), (110) and (111), but also to higher index planes.[4–9] It was found that chemical and thermal treatments can lead to the formation of surfaces in which one termination predominates over the others. In the (100) surface plane, for example, Nishimura *et al*. showed that a mainly $TiO_2$-termination can be obtained by a specific process.[5] Using chemical etching, Kobayashi *et al*. obtained an SrO termination in an $SrTiO_3$ (001) single crystal.[10] Recently, surface X-ray diffraction in ambient conditions has suggested that when the (100) surface is $TiO_2$ terminated there is most probably an oxygen overlayer.[11] The exact surface structure and chemistry will influence, e.g. for example, the work function and can significantly affect the electrical transport properties.[12] When considering polycrystalline ceramic samples the situation is even more complex, since we are in the presence of a multitude of different single-crystal orientations; each with the same lattice parameter and bulk stoichiometry. However, as Rahmati *et al*.[13] have demonstrated, the grain orientation can be a decisive factor in, e.g. Sr-rich island growth, and in the structure of these islands. Here we propose the first study on the surface chemical composition of polycrystalline Nb-doped STO, and its correlation with grain orientation.

The development of high spatial and energy resolution X-ray photoelectron emission microscopes (XPEEM) with high transmittance opens new possibilities in material analysis.[14,15] Chemical state sensitive elemental mapping can be performed with resolutions in the one hundred nanometer range directly at sample surfaces with high surface sensitivity (1 – 3 nm). This approach provides a good sample throughput associated to a non-time-consuming sample preparation. Also, spectroscopic images are easy to interpret and the energy resolution is that of high-resolution photoelectron spectroscopy, potentially one order of magnitude better than that typically obtained in electron energy loss transmission electron microscopy, thus making precise chemical state analysis possible with good lateral resolution. A novel XPEEM instrument has been recently commissioned at the nanocharacterisation Centre of the French Atomic Authority (CEA) in Grenoble and has already demonstrated promising capabilities.[16,17]

XPEEM analysis of polycrystalline niobium-doped STO was performed and correlated with electron backscattering diffraction (EBSD). Thanks to the high lateral resolution, each grain in the field of view of the instrument may be considered as a single-crystal sample having undergone precisely the same preparation procedure. In this way many single crystals were analysed in one single experiment with high energy resolution and high surface sensitivity. The results contribute to the understanding of the surface termination behaviour of STO.

## Experiment

Nb-doped $SrTiO_3$ samples were prepared from high purity strontium carbonate (99.9 + %, Sigma Aldrich), titania (99.9 + %, Sigma Aldrich) and niobium oxide (99.9%, ChemPur) powders.

* *Correspondence to: Nicholas Barrett, CEA-Saclay, DSM/IRAMIS/SPCSI, 91191 Gif-sur-Yvette, Cedex, France. E-mail: nick.barrett@cea.fr*

a *CEA-Saclay, DSM/IRAMIS/SPCSI, 91191 Gif-sur-Yvette, Cedex, France*

b *CEA-LETI, Minatec, 17 rue des Martyrs, 38054 Grenoble, Cedex 9, France*

c *Institut für Keramik im Maschinenbau, Universität Karlsruhe, D-76131 Karlsruhe, Germany*

d *Department of Materials, University of Oxford, Oxford OX1 3PH, UK*

After milling and calcination, the powder was uniaxially pressed into discs of 15 mm diameter and 5 mm thickness in a steel die and subsequently cold isostatically pressed at 400 MPa. The discs were heated to 1700 K at 20 K min$^{-1}$ in a 95% N$_2$ + 5% H$_2$ atmosphere and then kept at the temperature for 20 h. After sintering, the sample was quenched at more than 200 K min$^{-1}$. Further details on the synthesis procedure have been published elsewhere.[18] The sample diameter was reduced to 10.5 mm and then cut into discs of 1 mm thickness allowing the analysis of the central part. Finally the sample was polished down to 0.25 μm diamond paste (no chemical polishing was used). Once in the vacuum chamber (base pressure of 2 × 10$^{-8}$ Pa), the sample was annealed for 3 h at 1073 K. This annealing procedure has already been shown to yield a clean, ordered surface in single-crystal test samples oriented in the major directions.

The XPEEM instrument is a multi-mode, energy-filtered photoelectron microscope (NanoESCA, Omicron Nanotechnology GmbH) that can be operated both with laboratory and synchrotron X-rays.[15] Aberration-corrected energy filtering is performed with a double imaging hemispherical analyser which maintains high energy resolution and electron transmission at high spatial resolution.[14] Three operating modes of the microscope are available: direct PEEM, selected area spectroscopy, and energy-filtered imaging with both synchrotron radiation (SR) and laboratory (Al Kα) X-rays. The first mode is used for large field of views and alignments. High-resolution small-area spectroscopy is performed with the second mode, while the third is used for elemental and chemical state imaging with core-level electrons. The instrument was installed at the ID08 line of the European Synchrotron Radiation Facility (ESRF) which delivers soft X-rays over the 400 – 1500 eV photon energy range.[19] For spectroscopy, the photon energy was set to 665 eV and the total instrument resolution (beamline monochromator bandwidth and energy analyser) was 550 meV. For the core-level images, the photon energy was set to 665 eV for the O 1s and Ti 2p, and to 400 eV for the Sr 3d. The resulting photoelectron kinetic energies give an estimated inelastic mean free path (IMFP) of about 1 – 2 unit cells.[20,21] The lateral resolution estimated from a fit of the 16 – 84% intensity rise across a grain boundary is 0.25 μm.

After XPEEM experiments the same sample area was observed using scanning electron microscopy (SEM) and the grain orientation was determined by EBSD. High-quality Kikuchi patterns were easily obtained without any further sample preparation. The same sample was also analysed by high-resolution laboratory XPS using monochromatic Al Kα radiation at an angle of 60° with respect to the surface normal. These results provide a more bulk-like core-level standard whereas those obtained with XPEEM have a much higher surface sensitivity.

## Results and Discussion

Figure 1(a) shows the area-averaged core-level spectrum for O 1s obtained with a classical laboratory X-ray source, resulting in an IMFP of about four unit cells. The high-quality data demonstrate that there are probably two surface-related components corresponding to two distinct oxygen sites in the surface layer. In Fig. 1(b) the peak components obtained using SR and Al Kα are plotted, showing clearly the differences in the bulk-to-surface intensity ratios. However, there is no information on the grain-orientation dependence of the intensity ratios.

Figure 2(a) shows a threshold image taken at a photoelectron kinetic energy of 4.3 eV. A multitude of grains can be clearly observed with sizes varying from 1 to 5 μm. The contrast is attributed to variations in the work function according to grain orientation, although the absolute values are very sensitive to surface contamination.[17] Fig. 2(b) – (d) show images taken at O 1s, Sr 3d and Ti 2p core levels. The contrast in the core-level images can attain up to ±20% of the average signal, and to a first approximation is directly related to the elemental concentration. Given the mean free path for inelastic scattering, we can estimate the contribution from the surface layer to be at least 40% of the total core-level intensity. Thus the shallow information depth means that the contrast originates principally from variations in the chemical composition at the surface. The dark or grey colour indicates the lowest intensity observed in the image (after contrast/brightness adjustments). O, Sr and Ti spectra can be extracted from each grain in the image. The absolute intensity of the darkest grain is still far above the background at high kinetic energy. In Fig. 2 the arrow indicates a grain with high strontium intensity (concentration) and low titanium intensity (with respect to the average). This grain, with (100) orientation, seems to have a predominant SrO termination. The grain orientations were determined for all grains in the field of view by EBSD. In Fig. 3. we show the EBSD image recorded on the same sample area, the inset indicates the colour coding of the grain orientation. We can clearly resolve the three principal crystallographic directions and a series of intermediate orientations, confirming the multiplicity of crystalline orientations at the surface of the ceramic.

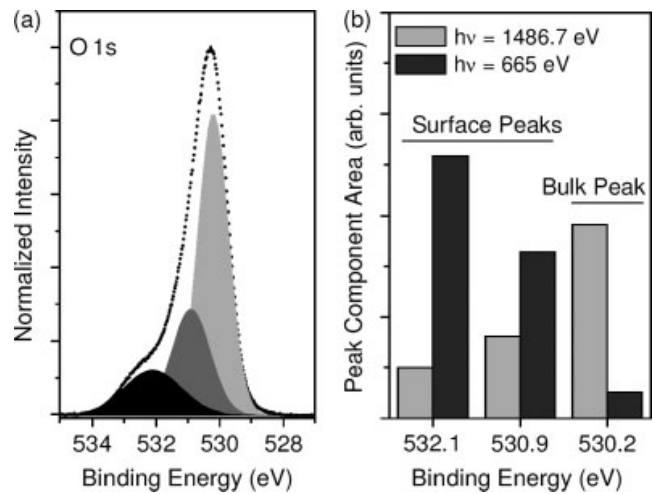

**Figure 1.** (a) O 1s spectrum obtained with Al Kα radiation deconvoluted in three peak components; (b) area of O 1s peak components for Al Kα and synchrotron radiation.

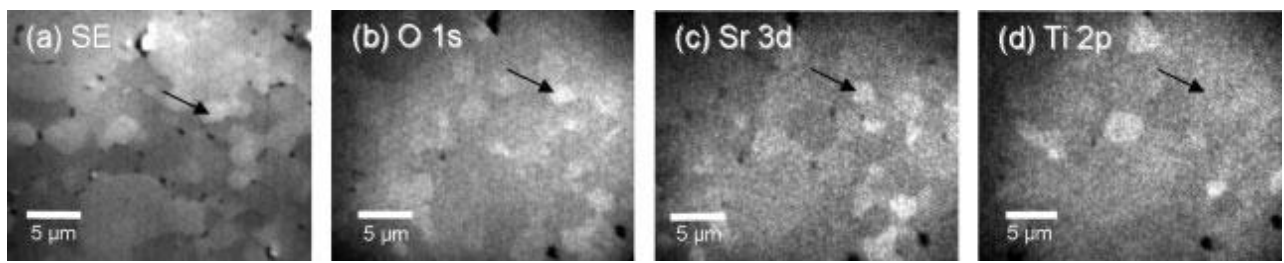

**Figure 2.** (a) Secondary electrons at a kinetic energy of 4.3 eV. Core-level images with respective binding energies: (b) O 1s, 530.5 eV; (c) Sr 3d, 266.5 eV; (d) Ti 2p, 206.0 eV. The arrow indicates a grain in the (100) direction.

By correlating the contrast in Fig. 2(b) – (d) with the grain orientations as determined from Fig. 3 we may plot the core-level signal intensities in a stereographic projection as a function of the grain orientation following the procedure already established by Rahmati *et al.*[13] Fig. 4(a) shows the principal directions in the stereographic plot. Figure 4(b) – (d) shows the stereographic intensity projections for the O 1s, Sr 3d and Ti 2p core levels. Over thirty grains were analysed, allowing an interpolated surface to be drawn covering all orientations. In Fig. 4 it is possible to observe, for instance, that near the (111) direction, Sr and O display lower signal intensity while Ti shows a high signal intensity. For this direction the termination layer would be preferentially Ti rich.

Finally we can extract spectra from single grains by plotting the photoemission intensity from the grain as a function of the kinetic in the image series recorded across a core level. Figure 5 presents spectra for three differently orientated grains extracted from a series of images across the O 1s core level, using the same partial components as for the area-averaged spectra in Fig. 1(a). Once again, the surface peaks are the major contribution. Furthermore, we can identify the origin of the contrast in the O 1s core-level image (Fig. 1(a)) as being the fingerprint of surface chemical states at lower binding energies, whereas the bulk (light grey) component has a small contribution independent of the grain orientation. It is shown that while the surface peak at higher binding energy (532.1 eV) is similar for all grain orientations, the other peak (at 530.9 eV) shows a very significant variation between grains. The O 1s emission at 532.1 eV is most probably due to a slight carbon contamination at the surface of the grains, and has already been observed.[10,22] The lowest binding energy peak at 530 eV is associated with bulk oxygen atoms.[10] Indeed, images taken close to 532 eV show no contrast or difference among different grains, indicating a (small) constant-surface carbon contribution but at the intermediate energy a clear contrast is observed as shown in Fig. 2(b). Therefore the component reflecting differences in the surface termination must be the one at 530.9 eV.

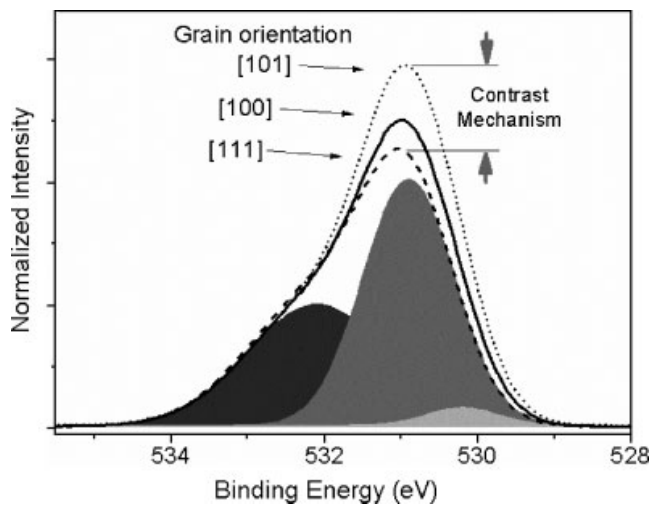

**Figure 5.** O 1s core level as determined from image series at different energies and using the same peak components as in Fig. 1. The peak components refer to the (100) direction.

## Conclusion

The high spatial and energy resolutions of a novel XPEEM instrument (NanoESCA) were employed in the study of the termination of Nb-doped SrTiO$_3$ surfaces. Individual grains within a polycrystalline material were imaged in the field of view of the microscope. This method permitted the analysis of over thirty different crystal orientations in one single experiment with the exactly same preparation procedure. The core-level images revealed an intensity contrast for different grains due to the elemental concentration at the grain surfaces. As a perovskite material, different surface terminations are expected for the $SrTiO_3$ and therefore different atomic concentrations in the last atomic layer. The intensity differences are well correlated with grain orientation as measured by EBSD. Stereographic plots of core-level intensities for different orientations are presented and average terminations can be attributed to each surface plane. High surface-sensitive spectra reveal indeed a significant difference with respect to more bulk-like high photon energy spectra, confirming the presence of surface-peak components. The O 1s spectra of individual grains show that the surface component at 530.9 eV is responsible for the grain orientation-dependent contrast observed in core-level images.

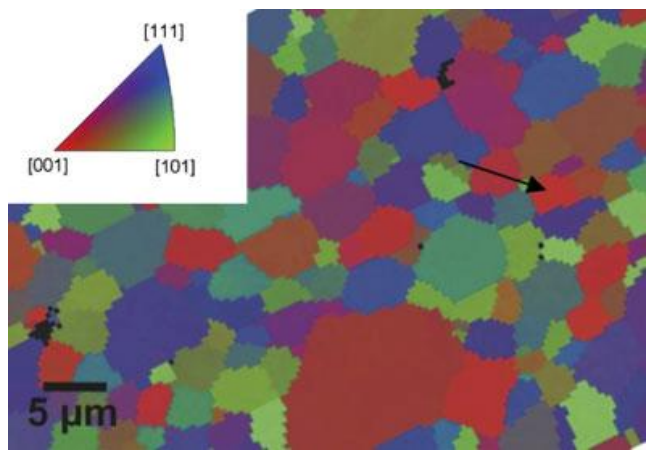

**Figure 3.** EBSD results indicate the grain orientation in the analysed area. The arrow indicates the same grain as in Fig. 2.

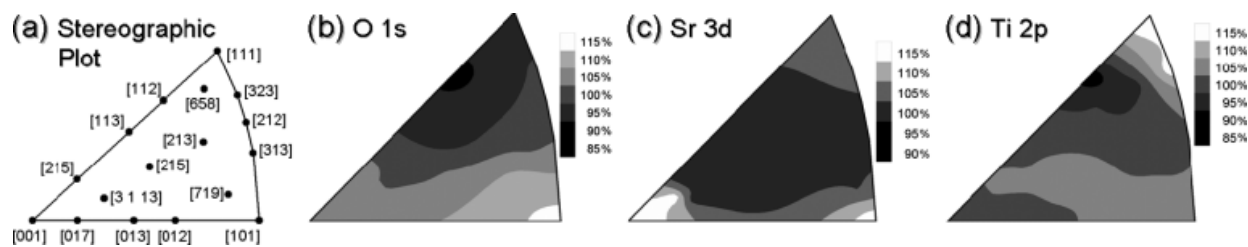

**Figure 4.** (a) Stereographic plot section indicating some crystal directions, (b) oxygen 1s, (c) strontium 3d, (d) titanium 2p core-level intensity variation as function of grain surface orientation plotted in a stereographic projection section. The scale refers to the variation with respect to the signal average of all grains considered.

**Acknowledgements**

We would like to acknowledge financial support by the European Commission under contract Nr. NMP3-CT-2005-013862 (INCEMS). We thank M. Gautier-Soyer for a critical reading of the manuscript.